\documentclass[a4paper]{amsart}
\usepackage{amssymb}

\newtheorem{theo}{Theorem}[section]

\newtheorem{lemm}[theo]{Lemma}
\newtheorem{alg}[theo]{Algorithm}

\newtheorem{defi}[theo]{Definition}
\theoremstyle{definition}
\newtheorem{ex}[theo]{Example}

\numberwithin{equation}{section}

\title[HSHP]{Hidden Sub-hypergroup Problem}
\author[Amini et al]{Massoud Amini , Mehrdad Kalantar, Mahmood M. Roozbehani}
\thanks{This research was in part supported by a grant from IPM (No. 84430017)}
\subjclass[2000]{68W40, 20N20} \keywords{hidden subgroup problem,
hypergroup, quantum fourier transform}
\address{Department of Mathematics, Tarbiat Modarres University, P.O.Box
14115-175, Tehran, Iran, \,\, and \newline Institute for Studies
in Theoretical Physics and Mathematics, Niavaran Square Tehran,
Iran}
\address{Faculty of Mathematical Sciences, Sharif University of Technology, Azadi Ave.,
Tehran, Iran}
\address{Department of Mathematics, Tarbiat Modarres University, P.O.Box
14115-175, Tehran, Iran}

\email{mamini@modares.ac.ir, kalantar@sharif.edu,
mmroozbehani@modares.ac.ir}

\begin{document}
\maketitle
\begin{abstract}
The Hidden Subgroup Problem is used in many quantum algorithms
such as Simon's algorithm and Shor's factoring and discrete log
algorithms. A polynomial time solution is known in case of abelian
groups, and normal subgroups of arbitrary finite groups. The
general case is still open. An efficient solution of the problem
for symmetric group $S_n$ would give rise to an efficient quantum
algorithm for Graph Isomorphism Problem. We formulate a hidden
sub-hypergroup problem for finite hypergroups and solve it for
finite commutative hypergroups. The given algorithm is efficient
if the corresponding QFT could be calculated efficiently.

\end{abstract}

\section{bacground}
Peter Shor in his seminal paper presented efficient quantum
algorithms for computing integer factorizations and discrete
logarithms. These algorithms are based on an efficient solution to
the hidden subgroup problem (HSP) for certain abelian groups. HSP
was already appeared in Simon's algorithm implicitly in form of
distinguishing the trivial subgroup from a subgroup of order 2 of
$\mathbb Z_{2^n}$.

The efficient algorithm for the abelian HSP uses the Fourier
transform. Other methods have been applied by Mosca and Ekert
[12]. The fastest currently known (quantum) algorithm for
computing the Fourier transform over abelian groups was given by
Hales and Hallgren [7]. Kitaev [10] has shown us how to
efficiently compute the Fourier transform over any abelian group
(see also [9]).

For general groups, Ettinger, Hoyer and Knill [5] have shown that
the HSP has polynomial query complexity, giving an algorithm that
makes an exponential number of measurements. Several specific
non-abelian HSP have been studied by Ettinger and Hoyer [4],
Rotteler and Beth [15], and Puschel, Rotteler, and Beth [14].
Ivanyos, Mangniez, and Santha [9] have shown how to reduce certain
non-abelian HSP's to an abelian HSP. The non-abelian HSP for
normal subgroups is solved by Hallgren, Russell, and Ta-Shma [8].

As for the Graph Isomorphism Problem (GIP), which is a special
case of HSP for the symmetric group $S_n$, Grigni, Schulman,
Vazirani and Vazirani [6] have independently shown that measuring
representations is not enough for solving GIP. However, they show
that the problem can be solved when the intersection of the
normalizers of all subgroups of G is large. Similar negative
results are obtained by Ettinger and Hoyer [4]. At the positive
side, Beals [3] showed how to efficiently compute the Fourier
transform over the symmetric group $S_n$ (see also [11]).

\begin{defi}(Hidden Subgroup Problem (HSP)). Given an
efficiently computable function $ f : G\to S$, from a finite group
G to a finite set S, that is constant on (left) cosets of some
subgroup H and takes distinct values on distinct cosets, determine
the subgroup H.
\end{defi}

An efficient quantum algorithms for abelian groups is as follows.

\begin{alg} (abelian HSP).

1. Prepare the state
$$\frac{1}{\sqrt {|G|}}\sum_{g\in G} |g\rangle |f (g)\rangle$$
and measure the second register, the resulting state is
$$\frac{1}{\sqrt {|H|}}\sum_{h\in H} |ch\rangle |f (ch) \rangle $$
where $c$ is an element of $G$ selected uniformly at random.

2. Compute the Fourier transform of the "coset" state above,
resulting in
$$\frac{1}{\sqrt{|H|.|G|}}\sum_{\rho\in\hat G}
\sum_{h\in H} \rho(ch)|\rho\rangle |f (ch) \rangle $$ where $\hat
G$ denotes the Pontryagin dual of $G$, namely the set of
homomorphisms $\rho : G\to \mathbb C$.

3. Measure the first register and observe a homomorphism $\rho$.
\end{alg}

Note that the resulting distribution over $\rho$ is independent of
the coset $cH$ arising after the first stage, as the support of
the first register in (1). Thus, repetitions of this experiment
result in the same distribution over $\hat G$. Also by the
principle of delayed measurement, measuring the second register in
the first step can in fact be delayed until the end of the
experiment.

\begin{alg}(non-abelian HSP, normal case)
1. Prepare the state $\sum_{g\in G} |g\rangle |f (g)\rangle$ and
measure the second register $|f (g) \rangle $. The resulting state
is $\sum_{h\in H} |ch\rangle |f (ch) \rangle $ where $c$ is an
element of $G$ selected uniformly at random. As above, this state
is supported on a left coset $cH$ of $H$.

2. Let $\hat G$ denote the set of irreducible representations of G
and, for each $\rho\in\hat G$, fix a basis for the space on which
$\rho$ acts. Let $d_\rho$ denote the dimension of $\rho$. Compute
the Fourier transform of the coset state, resulting in
$$\sum_{\rho\in\hat G}\sum_{1\leq i,j\leq d_\rho}
\frac{\sqrt {d_\rho}}{\sqrt{|H|.|G|}}\sum_{h\in H} \rho(ch)|\rho,
i,j\rangle |f (ch) \rangle $$

3. Measure the first register and observe a representation $\rho$.
\end{alg}

As before, one wishes the resulting distribution to be independent
of the actual coset $cH$ and depend only on the subgroup $H$. This
is guaranteed by measuring only the name of the representation
$\rho$ and leaving the matrix indices unobserved. The fact that
$O(log(|G|))$ samples of this distribution are enough to determine
$H$ with high probability is proved in [8].

\section{hypergroup representations}

A finite hypergroup is a set $K = \{c_0, c_1, \dots , c_n\}$
together with a $*$-algebra structure on the complex vector space
$\mathbb C K$ spanned by $K$ which satisfies the following axioms.
The product of elements is given by the structure equations
$$c_i*c_j =\sum_k n_{i,j}^k c_k,$$
with the convention that summations always range over $\{0, 1,
\dots , n\}$. The axioms are
\begin{enumerate}
\item $n_{i,j}^k\in\mathbb R$ and $n_{i,j}^k\geq 0,$

\item $\sum_k n_{i,j}^k= 1,$

\item $c_0*c_i=c_i*c_0=c_i,$

\item $K^*= K, \, n_{i,j}^0\neq 0$ if and only if $c^*_i = c_j,$
\end{enumerate}
for each $0\leq i,j,k\leq n$.

If $c^*_i = ci$, for each $i$, then the hypergroup is called
hermitian. If $c_i*c_j = c_j*c_i$, for each $i,j$, then the
hypergroup is called commutative. Hermitian hypergroups are
automatically commutative.

In harmonic analysis terminology, we have a convolution structure
on the measure algebra $M(K)$. This means that we can convolve
finitely additive measures on $K$ and, for $x,y\in K$, the
convolution $\delta_x*\delta_y$ is a probability measure. Indeed
$\delta_{c_i}*\delta_{c_j}\{c_k\}=n_{i,j}^k$. We follow the
convention of harmonic analysis texts and denote the involution by
$x\mapsto \bar x$ (instead of $x^*$), and the identity element by
$e$ (instead of $c_0$). For a function $f:K\to \mathbb C$, and
sets $A,B\subseteq K$ we put
$$f(x*y)=\sum_{z\in K} f(z)(\delta_x*\delta_y)\{z\},\quad (x,y\in K),$$
and
$$A*B=\cup\{supp(\delta_x*\delta_y): x\in A, y\in B\}.$$
A finite hypergroup $K$ always has a left Haar measure (positive,
left translation invariant, finitely additive measure)
$\omega=\omega_K$ given by
$$\omega\{x\}=\big((\delta_{\bar x}*\delta_x)\{e\}\big)^{-1}\quad
(x\in K).$$ A function $\rho:K\to \mathbb C$ is called a character
if $\rho(e)=1, \rho(x*y)=\rho(x)\rho(y)$, and $\rho(\bar
x)=\overline{\rho(x)}$. In contrast with the group case,
characters are not necessarily constant on conjugacy classes. Let
$K$ be a finite commutative hypergroup, then $\hat K$ denotes the
set of characters on $K$. In this case, for $\mu\in M(K)$ and
$f\in \ell^2(K)$, we put
$$\hat\mu(\rho)=\sum_{x\in K} {\rho(x)}\mu\{x\},\quad
\hat f(\rho)=\sum_{x\in K} f(x){\rho(x)}\omega\{x\}\quad
(\rho\in\hat K).$$ Hence $\hat f=(f\omega)\,\hat{}$. If
$H\subseteq K$ is a subhypergroup (i.e. $\bar H=H$ and
$H*H\subseteq H$), then $\hat\omega_H=\chi_{H^\perp}$ [2, 2.1.8],
where the right hand side is the indicator (characteristic)
function of
$$H^\perp=\{\rho\in\hat K: \rho(x)=1\,\, (x\in H)\}.$$
If $K/H$ is the coset hypergroup (which is the same as the double
coset hypergroup $K//H$ in finite case [2, 1.5.7]) with hypergroup
epimorphism (quotient map) $q:K\to H/K$ [2, 1.5.22], then
$(K/H)\,\hat{}\simeq H^\perp$ (with isomorphism map $\chi\mapsto
\chi\circ q$) [2, 2.2.26, 2.4.8]. Moreover, for each $\mu\in
M(K)$, $q(\mu *\omega_H)=q(\mu)$ [2, 1.5.12]. We say that $K$ is
strong if $\hat K$ is a hypergroup with respect to some
convolution satisfying
$$(\rho*\sigma)\,\check{}=\rho\,\check{}\sigma\,\check{}\quad(\rho,\sigma\in\hat
K),$$ where
$$\check k(x)=\sum_{\rho\in\hat K} k(\rho)\rho(x)\pi\{\rho\}\quad
(x\in K, k\in\ell^2(\hat K,\pi))$$ is the inverse Fourier
transform. In this case, for $\rho,\sigma\in\hat K$, we have
$\rho\in\sigma *H^\perp$ if and only if $Res_H\rho=Res_H\sigma$,
where $Res_H:\hat K\to\hat H$ is the restriction map [2, 2.4.15].
Also $H$ is strong and $\hat K/H^\perp\simeq \hat H$ [2, 2.4.16].
Moreover $(\hat K)\,\hat{}\simeq K$ [2, 2.4.18].

Let us quote the following theorem from [2, 2.2.13] which is the
cornerstone of the Fourier analysis on commutative hypergroups.

\begin{theo}{\bf (Levitan)}
If $K$ is a finite commutative hypergroup with Haar measure
$\omega$, there is a positive measure $\pi$ on $\hat K$ (called
the Plancherel measure) such that
$$\sum_{x\in K} |f(x)|^2\omega\{x\}=\sum_{\rho\in \hat K} |\hat
f(\rho)|^2\pi\{\rho\}\quad (f\in\ell^2(K,\omega)).$$ Moreover
$supp(\pi)=\hat K$ and $\pi\{\rho\}=\pi\{\bar\rho\}$. In
particular the Fourier transform $\mathfrak F$ is a unitary map
from $\ell^2(K,\omega)$ onto $\ell^2(\hat K,\pi)$.
\end{theo}

In quantum computation notation,
$$\mathfrak F: |x\rangle\mapsto\frac{1}{\tau(x)}\sum_{\rho\in\hat
K} \rho(x)\pi\{\rho\}|\rho\rangle,$$ where
$$\tau(x)=\big(\sum_{\rho\in \hat K}
|\rho(x)|^2\pi^2\{\rho\}\big)^{\frac{1}{2}}\quad(x\in K).$$ When
$K$ is a group, $\tau(x)=|\hat K|^{\frac{1}{2}}$, for each $x\in
K$. It is essential for quantum computation purposes to associate
a unitary matrix to each quantum gate. however, if we write the
matrix of $\mathfrak F$ naively using the above formula we don't
get a unitary matrix. The reason is that, in contrast with the
group case, the discrete measures on $\ell^2$ spaces are not
counting measure. More specifically, when $K$ is a group,
$\ell^2(K)=\bigoplus_{x\in K} \mathbb C$, where as here
$\ell^2(K,\omega)=\bigoplus_{x\in K}
\omega\{x\}^{\frac{1}{2}}\mathbb C$ and
$\ell^2(K)=\bigoplus_{\rho\in \hat K}
\pi\{\rho\}^{\frac{1}{2}}\mathbb C$. The exponent $\frac{1}{2}$ is
needed to get the same inner product on both sides. If we use
change of bases $|x\rangle^{'}=\omega\{x\}^{\frac{1}{2}}|x\rangle$
and $|\rho\rangle^{'}=\pi\{\rho\}^{\frac{1}{2}}|\rho\rangle$, the
Fourier transform can be written as
$$\mathfrak F: |x\rangle^{'}\mapsto \omega\{x\}^{\frac{1}{2}}\sum_{\rho\in\hat
K} \rho(\bar x)\pi\{\rho\}^{\frac{1}{2}}|\rho\rangle^{'},$$ and
the corresponding matrix turns out to be unitary.

There are not many finite hypergroups whose character table is
known [Wil]. Here we give two classical examples (of order two and
three and compute the corresponding Fourier matrix.

\begin{ex}
[Ross] The general form of an hypergroup of order 2 is known. It
is denoted by $K=\mathbb Z_2(\theta)$ and consists of two elements
$0$ and $1$ with multiplication table

\vspace{.5 cm}
\begin{center}
\begin{tabular}{|c|c|c|}
  \hline
  $*$ & $\delta_0$ & $\delta_1$ \\
  \hline
  $\delta_0$ & $\delta_0$ & $\delta_1$ \\
  \hline
  $\delta_1$ & $\delta_1$ & $\theta\delta_0+(1-\theta)\delta_1$ \\
  \hline
\end{tabular}
\end{center}
\vspace{.3 cm}

and Haar measure and character table

\vspace{.5 cm}
\begin{center}
\begin{tabular}{|c|c|c|}
  \hline
   & $0$ & $1$ \\
  \hline
  $\omega$ & $1$ & $\frac{1}{\theta}$ \\
  \hline
  $\chi_0$ & $1$ & $1$ \\
  \hline
  $\chi_1$ & $1$ & $-\theta$ \\
  \hline
\end{tabular}
\end{center}

\vspace{.3 cm} When $\theta=1$ we get $K=\mathbb Z_2$. The dual
hypergroup is again $\mathbb Z_2(\theta)$ with the plancherel
measure \vspace{.5 cm}
\begin{center}
\begin{tabular}{|c|c|c|}
  \hline
   & $\chi_0$ & $\chi_1$ \\
  \hline
  $\pi$ & $\frac{\theta}{1+\theta}$ & $\frac{1}{1+\theta}$ \\
  \hline
\end{tabular}
\end{center}
\vspace{.3 cm} The unitary matrix of the corresponding Fourier
transform is given by

$$\mathfrak F_2=\frac{1}{\sqrt{1+\theta^2}}\left(%
\begin{array}{cc}
  \theta &1 \\
  1 & -\theta \\
\end{array}%
\right)
$$
\end{ex}

\begin{ex}
[Wildberger] The general form of hypergroups of order 3 is also
known. We know that it is always commutative, but in this case,
the Hermitian and non Hermitian case should be treated separately.
Let $K=\{0,1,2\}$ be a Hermitian hypergroup of order three and put
$\omega_i=\omega\{i\}$, for $i=0,1,2$. Then the multiplication
table of $K$ is

\vspace{.5 cm}
\begin{center}
\begin{tabular}{|c|c|c|c|}
  \hline
  $*$ & $\delta_0$ & $\delta_1$ & $\delta_2$ \\
  \hline
  $\delta_0$ & $\delta_0$ & $\delta_1$ & $\delta_2$ \\
  \hline
  $\delta_1$ & $\delta_1$ & $\frac{1}{\omega_1}\delta_0+\alpha_1\delta_1+\beta_1\delta_2$
  & $\gamma_1\delta_1+\gamma_2\delta_2$ \\
  \hline
  $\delta_2$ & $\delta_2$ & $\gamma_1\delta_1+\gamma_2\delta_2$ & $\frac{1}{\omega_2}
  \delta_0+\beta_2\delta_1+\alpha_2\delta_2$ \\
  \hline
\end{tabular}
\end{center}
\vspace{.3 cm} where
$$\beta_1=\frac{\gamma_1\omega_2}{\omega_1},\,\, \beta_2=\frac{\gamma_2\omega_1}
{\omega_2},\,\,
\alpha_1=1-\frac{1+\gamma_1\omega_2} {\omega_1},\,\,
\alpha_2=1-\frac{1+\gamma_2\omega_1}{\omega_2}\,\,\gamma_2=1-\gamma_1,$$
and $\gamma_1$, $\omega_1$ and $\omega_2$ are arbitrary parameters
subject to conditions $0\leq \gamma_1\leq 1$, $\omega_1\geq 1$,
$\omega_2\geq 1$, and
$$1+\gamma_1\omega_2\leq \omega_1$$ $$1+(1-\gamma_1)\omega_1\leq \omega_2.$$
The Plancherel measure and character table
are given by

\vspace{.5 cm}
\begin{center}
\begin{tabular}{|c|c|c|c|c|}
  \hline
   & $\pi$ & $0$ & $1$ & $2$ \\
  \hline
   $\chi_0$ & $\frac{s_1}{t}$ & $1$ & $1$ & $1$\\
  \hline
  $\chi_1$ & $\frac{s_2}{t}$ & $1$ & $x$ & $z$ \\
  \hline
  $\chi_2$ & $\frac{s_3}{t}$ & $1$ & $y$ & $v$ \\
  \hline
\end{tabular}
\end{center}
 where
 $$x=\frac{\alpha_1-\gamma_1}{2}+\frac{D}{2\omega_2},\quad y=\frac{\alpha_1-\gamma_1}
 {2}-\frac{D}{2\omega_2} $$
 $$z=\frac{\alpha_2-\gamma_2}{2}-\frac{D}{2\omega_2},\quad v=\frac{\alpha_2-\gamma_2}{2}
 +\frac{D}{2\omega_2} $$
$$D=\sqrt{(1+\gamma_1\omega_2-\gamma_2\omega_1)^2+4\gamma_2\omega_1}$$ and
$$s_1=x^2v^2+\frac{y^2}{\omega_2}+\frac{z^2}{\omega_1}-(y^2z^2+\frac{x^2}{\omega_2}
+\frac{v^2}{\omega_1})$$
$$s_2=y^2+\frac{v^2}{\omega_1}+\frac{1}{\omega_2}-(v^2+\frac{y^2}{\omega_2}+\frac{1}
{\omega_1})$$
$$s_3=z^2+\frac{x^2}{\omega_2}+\frac{1}{\omega_1}-(x^2+\frac{z^2}{\omega_1}+\frac{1}
{\omega_1})$$
$$t=x^2v^2+y^2+z^2-(x^2+y^2z^2+v^2).$$
Let $\pi_i=\pi\{\chi_i\}=\frac{s_i}{t}$ and
$w_{ij}=\sqrt{\omega_i\pi_j}$, for $i,j=0,1,2$, then the Fourier
transform is given by the unitary matrix
$$\mathfrak F_3=\left(%
\begin{array}{ccc}
w_{00} & w_{10} & w_{20} \\
w_{01} & xw_{11} & zw_{21} \\
w_{02} & yw_{12} & vw_{22} \\
\end{array}%
\right)
$$
One concrete example is the normalized Bose Mesner algebra of the
square. In this case, $\omega_1=1, \omega_2=2,
\gamma_1=\beta_1=\alpha_1=\alpha_2=0, \gamma_2=1$, and
$\beta_2=\frac{1}{2}$. A simple calculation gives $D=2, x=1,
y=z=-1, v=0,$ and if we put $\pi_1=\frac{1}{4}$, we get
$\pi_2=\frac{1}{4}$ and $\pi_3=\frac{1}{2}$. In this case, the
Fourier transform matrix is
$$\mathfrak F_3=\frac{1}{2}\left(%
\begin{array}{ccc}
1 & 1 & \sqrt{2} \\
1 & 1 & -\sqrt{2} \\
\sqrt{2} & -\sqrt{2} & 0 \\
\end{array}%
\right)
$$

\vspace{.3 cm} In the non-Hermitian case, the multiplication table
of $K$ is

\vspace{.5 cm}
\begin{center}
\begin{tabular}{|c|c|c|c|}
  \hline
  $*$ & $\delta_0$ & $\delta_1$ & $\delta_2$ \\
  \hline
  $\delta_0$ & $\delta_0$ & $\delta_1$ & $\delta_2$ \\
  \hline
  $\delta_1$ & $\delta_1$ & $\gamma\delta_1+(1-\gamma)\delta_2$ &
  $\alpha\delta_0+\gamma\delta_1+\gamma\delta_2$ \\
  \hline
  $\delta_2$ & $\delta_2$ & $\alpha\delta_0+\gamma\delta_1+\gamma\delta_2$
  & $(1-\gamma)\delta_1+\gamma\delta_2$\\
  \hline
\end{tabular}
\end{center}
\vspace{.3 cm} where $\gamma=\frac{1-\alpha}{2}$, and $\alpha$ is
an arbitrary parameter with $0<\alpha\leq 1$. When $\alpha=1$, we
get $K=\mathbb Z_3$. The dual hypergroup is again $K$ and the
Plancherel measure and character table are given by \vspace{.5 cm}
\begin{center}
\begin{tabular}{|c|c|c|c|c|}
  \hline
   & $\pi$ & $0$ & $1$ & $2$ \\
  \hline
   $\chi_0$ & $\frac{s_1}{t}$ & $1$ & $1$ & $1$\\
  \hline
  $\chi_1$ & $\frac{s_2}{t}$ & $1$ & $z$ & $\bar z$ \\
  \hline
  $\chi_2$ & $\frac{s_2}{t}$ & $1$ & $\bar z$ & $z$ \\
  \hline
\end{tabular}
\end{center}
where $$z=\frac{-\alpha\pm i\sqrt{\alpha^2+2\alpha}}{2}.$$
$$s_1=2-\omega_1(\alpha^2+\alpha),\quad s_2=\omega_1-1,\quad
t=\omega_1(2-\alpha^2-\alpha).$$ Put $\pi_i=\pi\{\chi_i\}$ and
$w_{ij}=\sqrt{\omega_i\pi_j}$, for $i,j=0,1,2$, then the Fourier
transform is given by the unitary matrix
$$\mathfrak F_3=\left(%
\begin{array}{ccc}
w_{00} & w_{10} & w_{20} \\
w_{01} & zw_{11} & \bar zw_{21} \\
w_{02} & \bar zw_{12} & zw_{22} \\
\end{array}%
\right)
$$
As a concrete example, let us put $\omega_1=\omega_2=2,
\gamma=\frac{1}{4}$ and $\alpha=\frac{1}{2}$ to get
$z=\frac{-1+i\sqrt{5}}{4}$ and $\pi_1=\frac{1}{5}$,
$\pi_2=\pi_3=\frac{2}{5}$. In this case, the Fourier transform
matrix is
$$\mathfrak F_3=\frac{1}{\sqrt{5}}\left(%
\begin{array}{ccc}
1 & \sqrt{2} & \sqrt{2} \\
\sqrt{2} & \frac{-1+i\sqrt{5}}{4} & \frac{-1-i\sqrt{5}}{4} \\
\sqrt{2} & \frac{-1-i\sqrt{5}}{4} & \frac{-1+i\sqrt{5}}{4} \\
\end{array}%
\right)
$$
\end{ex}

\begin{lemm}
Let $K$ be commutative and $H$ be a sub-hypergroup of $K$ and
$\rho\in\hat K$, then the following are equivalent.

$(i)$ $\rho\in H^\perp$,

$(ii)$ $\sum_{m\in c*H} \omega\{m\}\rho(\bar m)\neq 0$, for each
$c\in K$,

$(iii)$ $\sum_{m\in c*H} \omega\{m\}\rho(m)\neq 0$, for some $c\in
K$.
\end{lemm}
\begin{proof} $(i)\Rightarrow (ii)$ If $\rho\in H^\perp$ and
$q:K\to K/H$ is the quotient map, then given $c\in K$,
$q(\mu*\omega_H)=q(\mu)$ for $\mu=\delta_c\omega\in M(K)$. But
clearly
$$q(\delta_c\omega)=\delta_{c*H}\omega=\sum_{m\in c*H} \delta_m\omega.$$
Hence $\rho(\delta_{c*H})\omega=\rho\circ q(\delta_c\omega)\neq
0$, where the last equality is because $\rho\circ q\in
(K/H)\,\hat{}$ and a character is never zero.

$(iii)\Rightarrow (i)$ If $\rho\notin H^\perp$ then the
multiplicative map $\rho\circ q$ should be identically zero on
$K/H$ (otherwise it is a character and $\rho\in H^\perp$). Hence
$\sum_{m\in c*H} \rho(m)\omega=\rho(\delta_{c*H})\omega= 0$, for
each $c\in K$.
\end{proof}

\section{HSHP}
In this section we give an algorithm for solving hidden
sub-hypergroup problem (HSHP) for abelian (strong) hypergroups.
This algorithm is efficient for those finite commutative
hypergroups whose Fourier transform is efficiently calculated. It
is desirable that, following Kitaev [10], one shows that the
Fourier transform could be efficiently calculated on each finite
commutative hypergroup. This could be difficult, as there is yet
no complete structure theory for finite commutative hypergroups
(see chapter 8 of [2]).

\begin{defi}(Hidden Sub-hypergroup Problem (HSHP)). Given an
efficiently computable function $ f : K\to S$, from a finite
hypergroup $K$ to a finite set $S$, that is constant on (left)
cosets of some subhypergroup $H$ and takes distinct values
$\lambda_c$ on distinct cosets $c*H$, for $c\in K$. Determine the
subhypergroup $H$.
\end{defi}

\begin{alg} (abelian HSHP).

1. Prepare the state $|\chi_0\rangle^{'}|0\rangle$.

2. Apply $\mathfrak F^{-1}$ to the first register to get
$$\sum_{x\in K} \omega\{x\}^{\frac{1}{2}}|x\rangle^{'}|0\rangle.$$

3. Apply the black box to get
$$\sum_{x\in K} \omega\{x\}^{\frac{1}{2}}|x\rangle^{'}|f(x)\rangle,$$
and measure the second register, to get
$$\frac{\sqrt{|K|}}{\sqrt{|c*H|}}\sum_{m\in c*H}
\omega\{m\}^{\frac{1}{2}}|m\rangle^{'} |\lambda_c \rangle,$$ where
$c$ is an element of $K$ selected uniformly at random, and
$\lambda_c$ is the value of $f$ on the coset $c*H$.

4. Apply $\mathfrak F$ to the first register to get
$$\frac{\sqrt{|K|}}{\sqrt{|c*H|}}\sum_{m\in c*H} \sum_{\rho\in\hat K}
\omega\{m\}\pi\{\rho\}^{\frac{1}{2}}\rho(m)|\rho\rangle^{'}
|\lambda_c
\rangle=\frac{\sqrt{|K|}}{\sqrt{|c*H|}}\sum_{\rho\in\hat
K}\pi\{\rho\}^{\frac{1}{2}}\big(\sum_{m\in
c*H}\omega\{m\}\rho(m)\big)|\rho\rangle^{'} |\lambda_c \rangle$$

5. Measure the first register and observe a character $\rho$.
\end{alg}

Note that the resulting distribution over $\rho$ is independent of
the coset $c*H$ arising after the first step. Also note that by
Lemma 2.2, the character observed in step 3 is in $H^\perp$.

\begin{theo} If the Fourier transform could be efficiently calculated on a
finite commutative hypergroup $K$, then the above algorithm solves
HSHP for $K$ in polynomial time.
\end{theo}

There ar a variety of examples of (commutative hypergroups) whose
dual object is known. One might hope to relate the HSP on a
(non-abelian) group $G$ to the HSHP on a corresponding commutative
hypergroup like $\hat G$ (see next example). The main difficulty
is to go from a function $f$ which is constant on cosets of some
subgroup $H\leq G$ to a function which is constant on cosets of a
subhypergroup of $\hat G$. The canonical candidate $\hat f$ fails
to be constant on costs of $H^\perp\leq \hat G$.

We list some of the examples of commutative hypergroups and their
duals, hoping that one can get such a relation in future.

\begin{ex}
If $G$ is a finite group, then $\hat G:=(G^G)\,\hat{}$ is a
commutative strong (and so Pontryagin [2, 2.4.18]) hypergroup [2,
8.1.43]. The dual hypergroups of the Dihedral group $D_n$ and the
(generalized) Quaternion group $Q_n$ are calculated in [2,
8.1.46,47].
\end{ex}

\begin{ex}
If $G$ is a finite group and $H$ is a (not necessarily normal)
subgroup of $G$ then the double coset space $G//H$ (which is
basically the same as the homogeneous space $G/H$ in the finite
case) is a hypergroup whose dual object is $A(\hat G, H)$ [2,
2.2.46]. It is easy to put conditions on $H$ so that $G//H$ is
commutative.
\end{ex}

There are also a vast class of special hypergroups (see chapter 3
of [2] for details) which are mainly infinite hypergroups, but one
might mimic the same constructions to get similar finite
hypergroups in some cases.

\end{document}